\documentstyle[12pt,epsf]{article}

\begin{document}


\begin{center}

 {\Large \bf
\vskip 7cm
\mbox{From the exclusive photoproduction of heavy quarkonia}
\mbox{at HERA to the EDDE}
\mbox{at TeVatron and LHC.}
}
\vskip 3cm

\mbox{Petrov~V.A.$^{a}$, Ryutin~R.A.$^{a}$}\\

\mbox{and}\\

\mbox{Prokudin~A.V.$^{a,b}$}

{\small\it
\vskip 0.2cm 
(a) Institute For High Energy Physics,\\ 
142281 Protvino,  RUSSIA}
\vskip 0.2cm 

{\small\it
\vskip 0.2cm 
(b)  Dipartimento di Fisica Teorica,\\ 
Universit\`a Degli Studi Di Torino, \\
Via Pietro Giuria 1,
10125 Torino, \\ 
ITALY\\
and\\
Sezione INFN di Torino,\\
 ITALY\\}

\vskip 1.75cm
{\bf
\mbox{Abstract}}
  \vskip 0.3cm

\newlength{\qqq}
\settowidth{\qqq}{In the framework of the operator product  expansion, the quark mass dependence of}
\hfill
\noindent
\begin{minipage}{\qqq}
Exclusive photoproduction of heavy quarkonia at HERA is analyzed in the framework of 
the Regge-eikonal approach together with the nonrelativistic bound state formalism. Total 
and differential cross-sections 
for the process $\gamma+p\to (Q\bar{Q})_{1S}+p$ are calculated. The model predicts
cross-sections of Exclusive Double Diffractive Events (EDDE) at TeVatron and LHC.

\end{minipage}
\end{center}


\begin{center}
\vskip 0.5cm
{\bf
\mbox{Keywords}}
\vskip 0.3cm

\settowidth{\qqq}{In the framework of the operator product  expansion, the quark mass dependence of}
\hfill
\noindent
\begin{minipage}{\qqq}
Exclusive photoproduction of vector mesons -- Pomeron -- Regge-Eikonal model
\end{minipage}

\end{center}

\setcounter{page}{1}
\newpage


\section{Introduction}

The study of properties of bound states of heavy quarks plays a central role 
in the understanding of strong interactions and verification of different QCD inspired
and nonperturbative models, because such processes give a very exciting possibility
to observe interplay of "hard" and "soft" regimes~\cite{hardnsoft}. There have been intensive experimental studies of the $J/\Psi$
and $\Upsilon$ production in $ep$ collisions at HERA~\cite{HERAexp1},\cite{HERAexp2} and also a lot of theoretical
investigations~\cite{QQth1}-\cite{QQth5}.

In this paper we consider exclusive photoproduction of $V=Q\bar{Q}_{1S}$ states from
another viewpoint. There are some other interesting processes that will be investigated at present and
future hadronic colliders. We need estimations of cross-sections for such processes and can use the data
from HERA as a source of normalization of phenomenological models. Here we show 
that the extended Regge-Eikonal approach~\cite{Petrov1}-\cite{3Pom} gives not only a good description of the data on exclusive 
vector meson photoproduction but can be used also to predict rates of Exclusive Double Diffractive 
Events (EDDE) at LHC and Tevatron. The advantages of these events have been considered in~\cite{EDDH},\cite{KMR}. 

\section{Calculations}

In Fig.~1 we illustrate in detail the process
$\gamma(q)+p(p)\to V(p_v)+p(p^{\prime})$. Off-shell proton-gluon amplitude $T$ in Fig.~1 is treated
by the method developed in Ref.~\cite{Petrov1}, which is based on the extension of Regge-eikonal 
approach, and succesfully used for the description of the data from hadron colliders~\cite{Petrov2}-\cite{3Pom}. The 
amplitude $A$ of the process $\gamma(q)+g(\kappa_1)\to V(p_v)+g(\kappa_2)$ is calculated in the nonrelativistic bound state 
approximation(see~\cite{QQth1}-\cite{QQth3} and ref. therein):

\begin{eqnarray}
A&=&\frac{R_{v0}}{\sqrt{16\pi M_v}}Sp\left[\hat{\cal{O}}(\hat{p}_v-M_v)\hat{\epsilon}_v\right]\\
\hat{\cal{O}}&=&e e_Q g^2\frac{\delta^{ab}}{2\sqrt{3}}\left( \frac{({p_{v}}_{\alpha}-2{\kappa_1}_{\alpha}+\hat{\kappa}_1\gamma_{\alpha})\hat{\epsilon}_{\gamma}({p_{v}}_{\beta}+2{\kappa_2}_{\beta}-\gamma_{\beta}\hat{\kappa}_2)}{(-p_v\kappa_1+\kappa_1^2+i0)(p_v\kappa_2+\kappa_2^2+i0)}+ 5\; permut.\right)\;,\;
\end{eqnarray}
where $p_v^2=M_v^2$, $e_Q$ is the charge of heavy quark, $R_{v0}$ is the absolute value of the
vector meson radial wave function at the origin, $\epsilon_{v,\gamma}$ are photon and vector meson 
polarization vectors correspondingly. Permutations are taken for all gauge 
bosons. Notations and vector decompositions that are used in the article are

\begin{eqnarray}
\label{nota}
&&\kappa_1=\kappa+\frac{\Delta}{2}\;,\; \kappa_2=\kappa-\frac{\Delta}{2}\;,\;p=p'+\frac{m_p^2}{s}q'\;,\;q=q'-\frac{Q^2}{s}p'\;,\;\\
&&Q^2=-q^2\;,\; p^2=m_p^2\;,\; q'^2=p'^2=0\;,\; s\simeq 2p'q'\;,\nonumber\\
&&\kappa=\frac{x_v}{2}(\alpha p'+\beta q')+\kappa_{\perp}\;,\;x_v=\frac{M_v^2}{s}\;,\; \kappa_{\perp}^2=-\vec{\kappa}^2\;,\; y=\frac{4\vec{\kappa}^2}{M_v^2}\;,\; y'=-\frac{4\kappa^2}{M_v^2}\;,\nonumber\\
&&\Delta=x_v\left(\left[ 1+y_Q+y_{\Delta}\right]p'-y_{\Delta}q'\right)+\Delta_{\perp}\;,\; y_Q=\frac{Q^2}{M_v^2}\;,\; t\simeq\Delta_{\perp}^2=-\vec{\Delta}^2\;,\; y_{\Delta}=\frac{\vec{\Delta}^2}{M_v^2}\;,\nonumber\\
&&p_v=q+\Delta\;,\; y_0=\frac{m_p^2}{M_v^2}\;\nonumber
\end{eqnarray}
Photon and vector meson polarization vectors in the general case ($Q\neq0$) can be represented as follows:
\begin{eqnarray}
\label{es}
&&{\epsilon_{\gamma}}_{\perp}q={\epsilon_{\gamma}}_0q=0\;,\;{\epsilon_{\gamma}}_{\perp}^2=-{\epsilon_{\gamma}}_0^2=-1\;,\;{\epsilon_{\gamma}}_0=\frac{1}{Q}(q'+x_v y_Q p')\;,\\
&&{\epsilon_v}_{\perp}p_v={\epsilon_v}_{\parallel}p_v=0\;,\;{\epsilon_v}_{\perp}=v_{\perp}+\frac{2(\vec{v}\vec{\Delta})}{s}(p'-q')\;,\;v_{\perp}^2=-\vec{v}^2\;,\;\nonumber\\
&&{\epsilon_v}_{\parallel}=\frac{1}{M_v}(q'-x_v(1-y_{\Delta})p'+\Delta_{\perp})\nonumber
\end{eqnarray}

For the amplitude of the process $\gamma(q)+p(p)\to V(p_v)+p(p^{\prime})$ we have:
\begin{equation}
\label{M1}
M=\int\frac{d^4\kappa}{(2\pi)^4}\frac{1}{(\kappa_1^2+i0)(\kappa_2^2+i0)}A^{\alpha\beta,\; a b}
T_{\alpha\beta,\; a b}
\end{equation}
\begin{eqnarray}
\label{Tamp}
&&T_{\alpha\beta,\; a b}=\delta_{a b} \left(G_{\alpha\beta}-\frac{{P_1}_{\alpha}{P_2}_{\beta}}{P_1P_2}\right) T^D_{gp\to gp}\;,\\
&&G_{\alpha\beta}=g_{\alpha\beta}-\frac{{\kappa_2}_{\alpha}{\kappa_1}_{\beta}}{\kappa_1\kappa_2}\;,\\
&&{P_1}=p-\frac{p\kappa_{1}}{\kappa_1\kappa_2}\kappa_{2}\;,\;{P_2}=p-\frac{p\kappa_{2}}{\kappa_1\kappa_2}\kappa_{1}\;,\;
\end{eqnarray}
 Generally the amplitude $T^D_{gp\to gp}$ can be represented in the
Regge-eikonal form~\cite{Petrov2},\cite{3Pom} with fixed parameters 
of trajectories from Ref.~\cite{3Pom} (see Table.~1), in
which  the eikonal is dominated by three vacuum trajectories (Pomerons with
different properties). It follows from the analysis below that at small $t$ 
the amplitude $T^D_{gp\to gp}$ takes the simple Regge form, which is dominated
by the 3rd ("hard") Pomeron:
\begin{equation}
\label{TReg}
T^D_{gp\to gp}\simeq c_{gp}\left(e^{-i\frac{\pi}{2}}\frac{2p\kappa}{s_0-\kappa^2}\right)^{\alpha_{P_3}(t)}e^{b^{(3)}_0 t}\;, b^{(3)}_0=\frac{r^2_{gP_3}+0.5 r^2_{pP_3}}{4}\;,
\end{equation}
where $s_0\simeq 1\; GeV$ is the scale parameter of the model that is used in the global
fitting of the data on $pp(p\bar{p})$ scattering~\cite{Petrov3},\cite{3Pom}, 
$r^2_{pP_3}$, $\alpha_{P_3}(t)=\alpha_{P_3}(0)+\alpha'_{P_3}(0) t$ are 
defined in Table.1,  
$r^2_{gP_3}$ and $c_{gp}$ are extracted by the procedure~(\ref{REMTT})-(\ref{JPsicgp}). With
notations (\ref{nota}) we have: 

\begin{equation}
\label{d4kappa}
d^4\kappa=\pi\frac{M_v^4 x_v}{32}d\alpha d\beta dy=-\pi\frac{M_v^4 x_v}{64}\frac{d\alpha}{\alpha} dy' dy
\end{equation}

In the limit $Q\to 0\;,\; t\to 0$ only the amplitude $M_{\perp\perp}$ survives:

\begin{eqnarray}
\label{MTT}
|M_{\perp\perp}|^2&\simeq& K_v^2 I_v(t)^2 c_{gp}^2 \left(\frac{s}{s_0}\right)^{2 \alpha_{P_3}(0)} e^{2 b_3 t}\;,\\
\label{b3def} b_3&=&b^{(3)}_0+\alpha'_{P_3}(0)\ln\frac{s}{s_0}\\
K_v^2&=&\frac{4096 \alpha_e \alpha_s^2 e_Q^2 |{R_v}_0|^2}{3 M_v^3 \pi^4}=\frac{1024\alpha_s^2\Gamma(V\to e^+e^-) K_{NLO}}{3 M_v\pi^4\alpha_e}\;,\\
I_v(t)&=&\int d\alpha dy'\int_0^1 dy\frac{f(\alpha,y,y')}{(\alpha-1-y'+i0)(\alpha+1+y'-i0)}\cdot\\
&\cdot&\frac{1}{(\alpha y'-y+y'-i0)(\alpha y'+y-y'-i0)}\nonumber\\
f(\alpha,y,y')&=&\frac{1}{2\alpha^{\alpha_{P_3}(t)-1}}\left[ \frac{\alpha^2 y y'}{(y-y')^2}\right]\left( \frac{y-y'}{2\left( 1+\frac{y'}{4y_0}\right)}\right)^{\alpha_{P_3}(t)}
\end{eqnarray}

Now let us extract the values of parameters from the fit to the data on elastic $J/\Psi$ photoproduction~\cite{HERAexp1}. At first we write
the amplitude $M_{\perp\perp}$ in the Regge-eikonal form with parameters 
from Table.1 and the coefficient that corresponds
to the simple Vector Dominance Model (VDM):

\begin{eqnarray}
\label{REMTT}
M_{\perp\perp}=\sqrt{\frac{3\Gamma(V\to e^+e^-)}{\alpha_e M_v}}\;4\pi s\int_0^\infty db^2 J_0(b\sqrt{-t})\frac{e^{2i(\delta_1+\delta_2+\delta_3)}-1}{2i}\;,
\end{eqnarray}
where
\begin{eqnarray}
\label{eikonals}
&&\delta_i=i\frac{c^{(i)}_{vp}}{s_0}\left( e^{-i\frac{\pi}{2}}\frac{s}{s_0}\right)^{\alpha_{P_i}(0)-1}\frac{e^{\frac{-b^2}{\rho_i^2}}}{4\pi\rho_i^2}\;,\\
&&\rho_i^2=4\alpha'_{P_i}(0)\ln\left( e^{-i\frac{\pi}{2}}\frac{s}{s_0}\right)+r^2_{gP_i}+0.5 r^2_{pP_i}\nonumber
\end{eqnarray}
As will be seen below, in our case the VDM plus Regge-eikonal approach representation~(\ref{REMTT}) 
is applicable.

Results of this fit for $J/\Psi$ meson are shown in 
Figs.2-5. As we see from figures, the main contribution to the cross-section is given by the Born 
term of the 3rd Pomeron. The 1st "soft" Pomeron gives no contribution. The term corresponding to the 2nd 
Pomeron vanishes faster with $t$, and gives the contribution less 
than 1\%, when $t\le -0.2\;GeV^2$. Numerical estimations show that 
absorbtive corrections play minor role at $t\simeq t^*=$\linebreak
$=-1/2b_3$, where $b_3$ is obtained from~(\ref{b3def}). Using these facts,
we keep in~(\ref{REMTT}) only the Born term for the 3rd 
Pomeron with parameters
\begin{equation}
\label{3Ppars}
r^2_{gP_3}= 2.54\pm 0.41\mbox{GeV}^{-2}\;,\; c^{(3)}_{J/\Psi p}=1.11\pm 0.07\;,\; \chi^2/dof=1.48\;
\end{equation}
and take the integral $I_v$ at $t=t^*$. Now we can estimate the constant $c_{gp}$ in~(\ref{MTT})
from the comparison of two formulae for the amplitude $M_{\perp\perp}$:

\begin{equation}
\label{cgp}
c_{gp}=\frac{\sqrt{\frac{3\Gamma(V\to e^+e^-)}{\alpha_e M_v}}}{K_v I_v(t^*)}\; c^{(3)}_{vp}=\frac{3 \pi^2}{32 \alpha_s I_v(t^*)\sqrt{K_{NLO}}}\; c^{(3)}_{vp}\;,\\
\end{equation}

 Taking for $J/\Psi$ mesons 
\begin{eqnarray}
\label{JPsipars}
&&M_{J/\Psi}=3.1\;\mbox{GeV}\;,\; \alpha_s(M_{J/\Psi}^2)=0.25\;,\;\\
&&I_{J/\Psi}(t^*)\simeq 0.83\;,\; 35\;\mbox{GeV}<W=\sqrt{s}<260\;\mbox{GeV}\;,\nonumber\\
&&\Gamma(J/\Psi\to e^+e^-)=5.26\pm 0.37\;\mbox{keV}\;,\; K_{NLO}\simeq 2\; \mbox{(see, for example, \cite{QQNLO})}\nonumber
\end{eqnarray}
we get from~(\ref{cgp}):

\begin{equation}
\label{JPsicgp}
c_{gp}=3.5\pm 0.4
\end{equation}
Here errors are estimated from uncertanties of quantities in~(\ref{cgp}).


The data on $\Upsilon$ production~\cite{HERAexp2} gives the possibility to check the model 
predictions. The result of ZEUS collaboration for the ratio of total cross-sections of $J/\Psi$ and
$\Upsilon$ photoproduction:

\begin{equation}
\label{ZEUSratio}
\frac{\sigma_{\gamma p\to\Upsilon p}}{\sigma_{\gamma p\to J/\Psi p}}=(4.8\pm 2.2(\mbox{stat.}){+0.7\atop -0.6} (\mbox{sys.}))\cdot 10^{-3}
\end{equation}
If we assume that the constant $c_{gp}$ is the same for both processes, and the slope of the exponent
does not change much with energy, then from the expression~(\ref{MTT}) we will get at the same value of $W$:

\begin{equation}
\label{teorratio}
\frac{\sigma_{\gamma p\to\Upsilon p}}{\sigma_{\gamma p\to J/\Psi p}}\simeq\left[ \frac{\alpha_s(M^2_{\Upsilon})I_{\Upsilon}}{\alpha_s(M^2_{J/\Psi})I_{J/\Psi}}\right]^2\frac{\Gamma(\Upsilon\to e^+e^-)K^{\Upsilon}_{NLO} M_{J/\Psi}}{\Gamma(J/\Psi\to e^+e^-)K^{J/\Psi}_{NLO} M_{\Upsilon}}
=(3.1\pm 1.1)\cdot 10^{-3}\;,
\end{equation}
where   
\begin{eqnarray}
\label{Kfactor}
&&\Gamma(\Upsilon\to e^+e^-)=1.32\pm 0.04\pm 0.03\;\mbox{keV}\;,\\
&&M_{\Upsilon}=9.46\;\mbox{GeV}\;,\;\alpha_s(M_{\Upsilon}^2)\simeq 0.2\;,\; I_{\Upsilon}\simeq 0.21\;,\nonumber\\
&&K_{NLO}\sim\frac{1}{1-\frac{16\alpha_s}{3\pi}}\; \mbox{(see Ref.~\cite{QQNLO})}\;,\nonumber
\end{eqnarray}
and uncertainty of the result originates from the errors of parameters in~(\ref{teorratio}). Theoretical
estimation does not contradict the experimental value~(\ref{ZEUSratio}).

 
The second estimation can be done for the EDD dijet production at TeVatron energies. Recent CDF
results~\cite{CDF1},\cite{CDF2} for the upper bound of the cross-section of the process 
$p+p\to p+jet+jet+p$ are the following:

\begin{eqnarray}
\label{dijetres}
&&E_T>7\;\mbox{GeV}\;,\; \sigma<3.7\; \mbox{nb}\;,\\
&&E_T>10\;\mbox{GeV}\;,\; \sigma<0.97\pm 0.065\; (\mbox{stat.})\pm 0.272 (\mbox{sys.})\; \mbox{nb}\;,\nonumber\\
&&E_T>25\;\mbox{GeV}\;,\; \sigma<34\pm 5\; (\mbox{stat.})\pm 10 (\mbox{sys.})\; \mbox{pb}\;.\nonumber
\end{eqnarray}
After theoretical calculations by the method developed in Refs.~\cite{EDDH},\cite{EDDE} we  
extract upper bounds for the parameter $c_{gp}$ from~(\ref{dijetres}): 

\begin{eqnarray}
\label{cgpCDF}
&&E_T>7\;\mbox{GeV}\;,\; c_{gp}<3.3\;,\\
&&E_T>10\;\mbox{GeV}\;,\; c_{gp}<3.4\;,\nonumber\\
&&E_T>25\;\mbox{GeV}\;,\; c_{gp}<4.2\;.\nonumber
\end{eqnarray}
Values of $c_{gp}$ are close to our estimation~(\ref{JPsicgp}).

\section*{Conclusions}

 We can conclude that the generalized Regge-eikonal approach with 3 different Pomerons describes well
the data on $J/\Psi$ production. The main contribution to the cross-section comes from the
term corresponding to the 3rd, so called, "hard" Pomeron. This makes possible
to extract the corresponding parameter of the model.
 
 The upper bound for the same parameter is found to be close
to our result, when calculated 
from experimental estimations on EDD di-jet production made by CDF. It indicates once more the applicability of the Regge-eikonal approach and
gives us the tool for further predictions.
 
\section*{Aknowledgements}

 This work is supported by the Russian Foundation for Basic Research, grant no. 04-02-17299 and partially by grant CNRS-PICS-2910.




\newpage

Table 1.: Parameters $\alpha_{P_i}(0)$, $\alpha'_{P_i}(0)$, $r^2_{pP_i}$ are obtained
from the fit to the data on $p(\bar{p})+p\to p(\bar{p})+p$~\cite{3Pom} and remain fixed during the $J/\Psi$
data fitting.

\vspace*{0.4cm}
\begin{tabular}{|c|c|c|c|}
\hline
          &         &    &   \\
 Pomeron$_i$        &      1   & 2   & 3  \\
          &         &    &   \\
\hline 		  
          &         &    &   \\
 $\alpha_{P_i}(0)-1$ &     $0.0578\pm 0.0020$     &   $0.1669\pm 0.0012$  &   $0.2032\pm 0.0041$ \\
          &         &    &   \\
\hline 
          &         &    &   \\
 $\alpha'_{P_i}(0)$ (GeV$^{-2}$)&    $0.5596\pm 0.0078$   &   $0.2733\pm 0.0056$  &   $0.0937\pm 0.0029$ \\
          &         &    &   \\
\hline 
          &         &    &   \\
  $r^2_{pP_i}$ (GeV$^{-2}$)&     $6.3096\pm 0.2522$   &    $3.1097\pm 0.1817$ &   $2.4771\pm 0.0964$\\
          &         &    &   \\
\hline 
\end{tabular}
\vspace*{0.4cm}

\newpage
\section*{Figure captions}

\begin{list}{Fig.}{}

\item 1: Diagram for the process $\gamma+p\to V+p$.
\item 2: Diagrams for the process $\gamma+g^*\to V+g^*$.
\item 3-5: Differential cross-sections of the process $\gamma+p\to V+p$ at different values of W. Solid line is the Born term for the 3rd Pomeron and dashed curve is the unitarized result.
\item 6: Total cross-section of the process $\gamma+p\to V+p$. Solid line is the Born term for the 3rd Pomeron and dashed curve is the unitarized result.

\end{list}


\newpage

\begin{figure}[h]
\vskip 1cm
\hskip  1cm \vbox to 7cm {\hbox to 7cm{\epsfxsize=7cm\epsfysize=7cm\epsffile{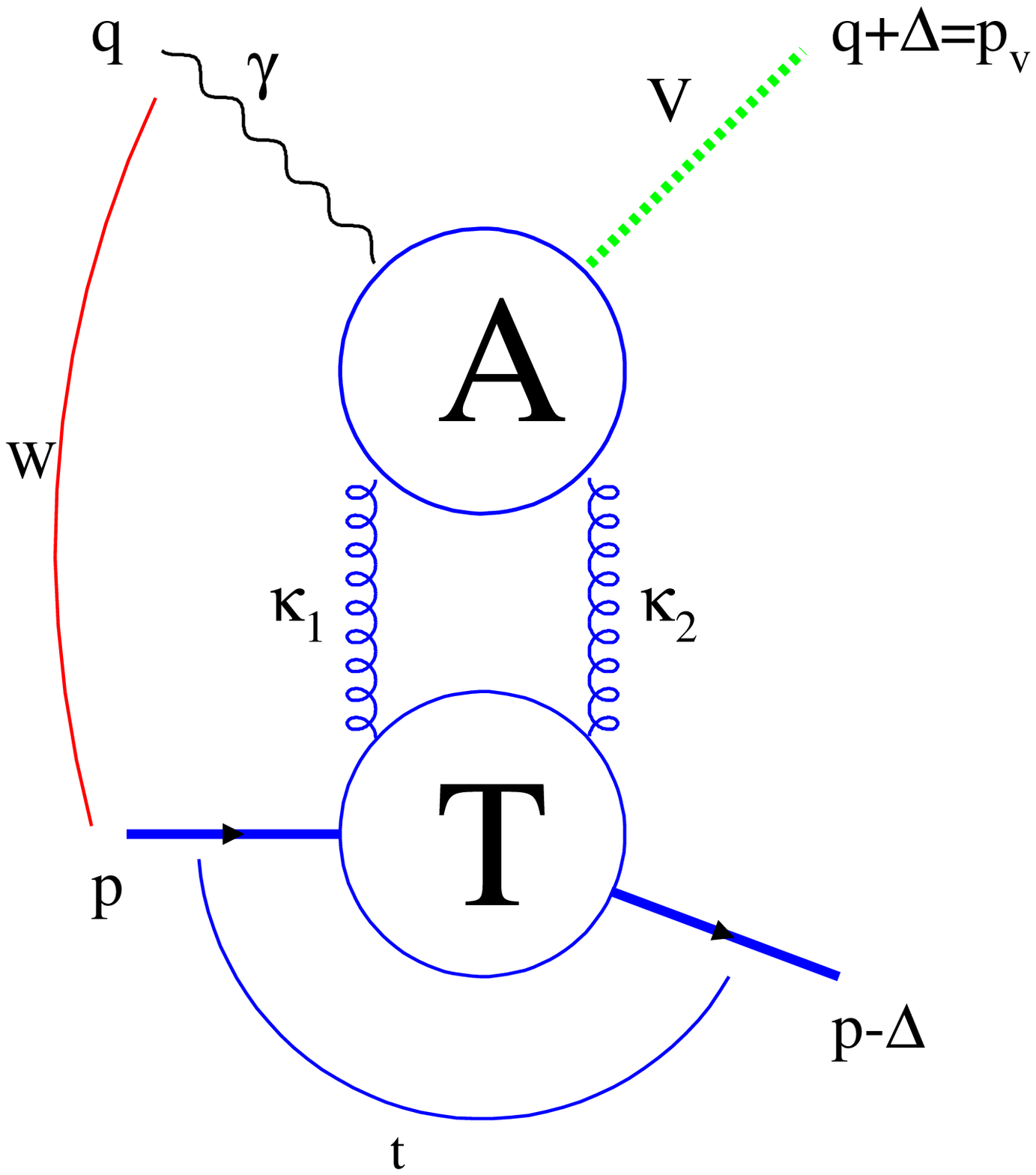}}}
\hskip 1cm
\caption{}
\end{figure}

\begin{figure}[h]
\vskip 1cm
\hskip  1cm \vbox to 5cm {\hbox to 16cm{\epsfxsize=16cm\epsfysize=5cm\epsffile{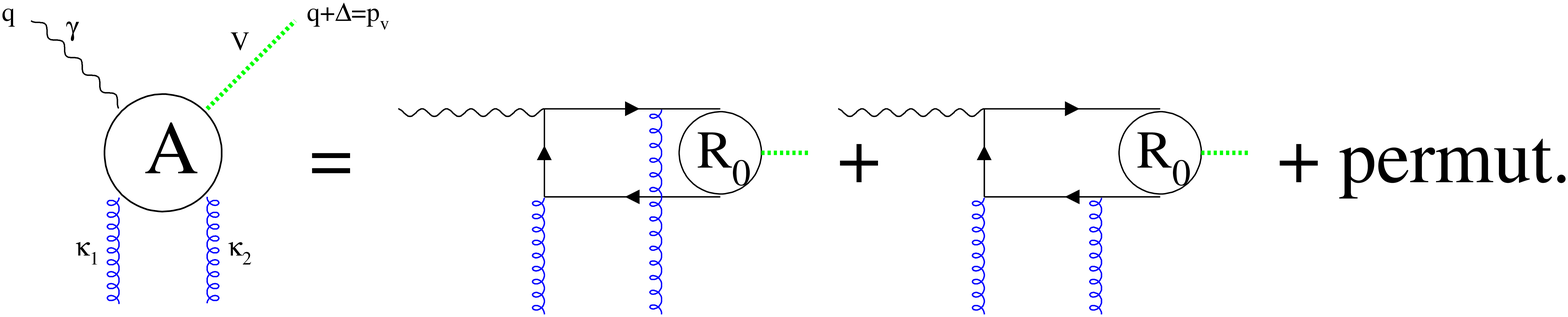}}}
\hskip 1cm
\caption{}
\end{figure}

\newpage

\begin{figure}[hb]
\vskip 1.5cm
\vbox to 18cm {\hbox to 16cm{\epsfxsize=16cm\epsfysize=18cm\epsffile{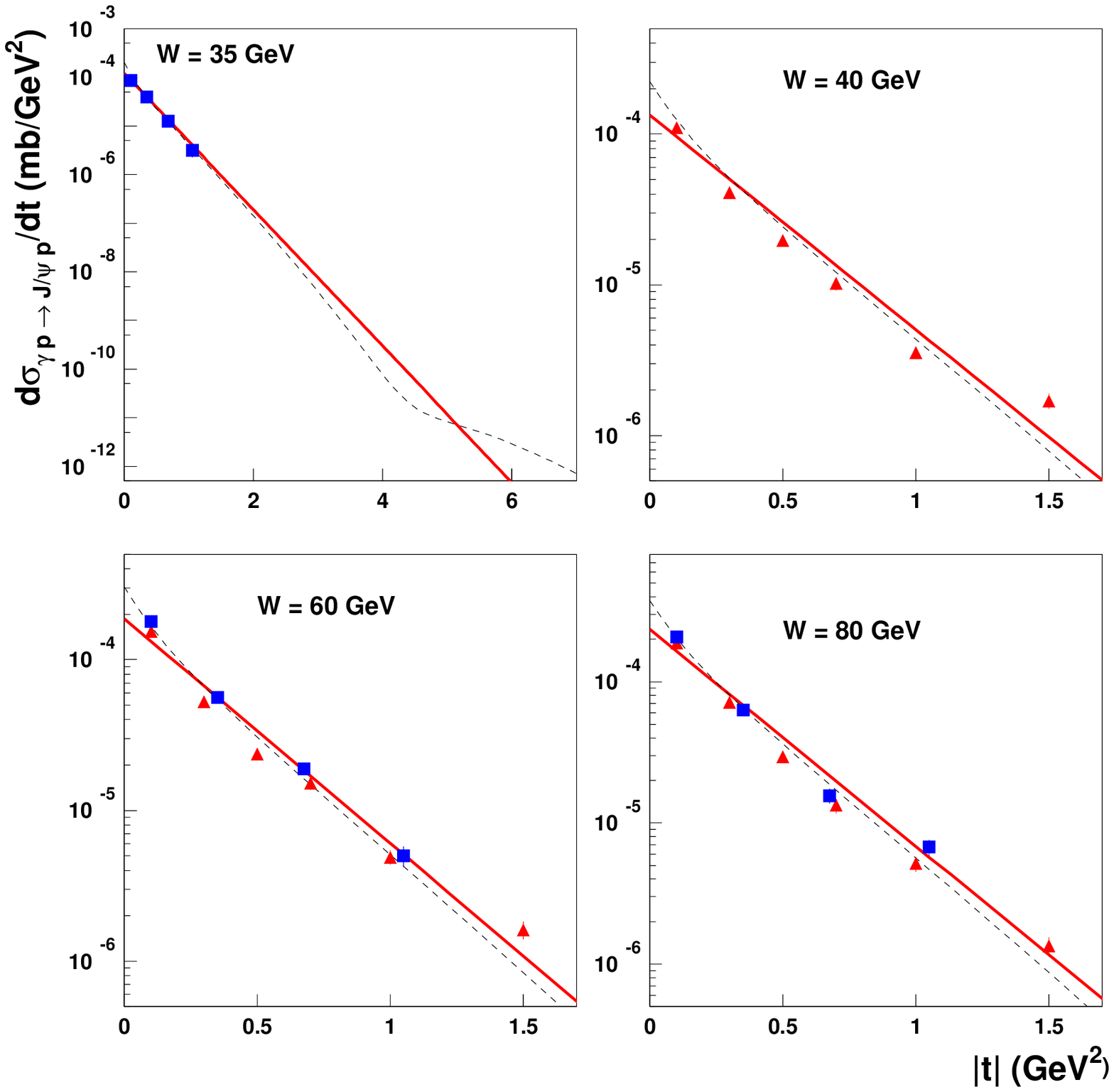}}}
\hskip 1cm
\caption{}
\end{figure}

\newpage

\begin{figure}[hb]
\vskip 1.5cm
\vbox to 18cm {\hbox to 16cm{\epsfxsize=16cm\epsfysize=18cm\epsffile{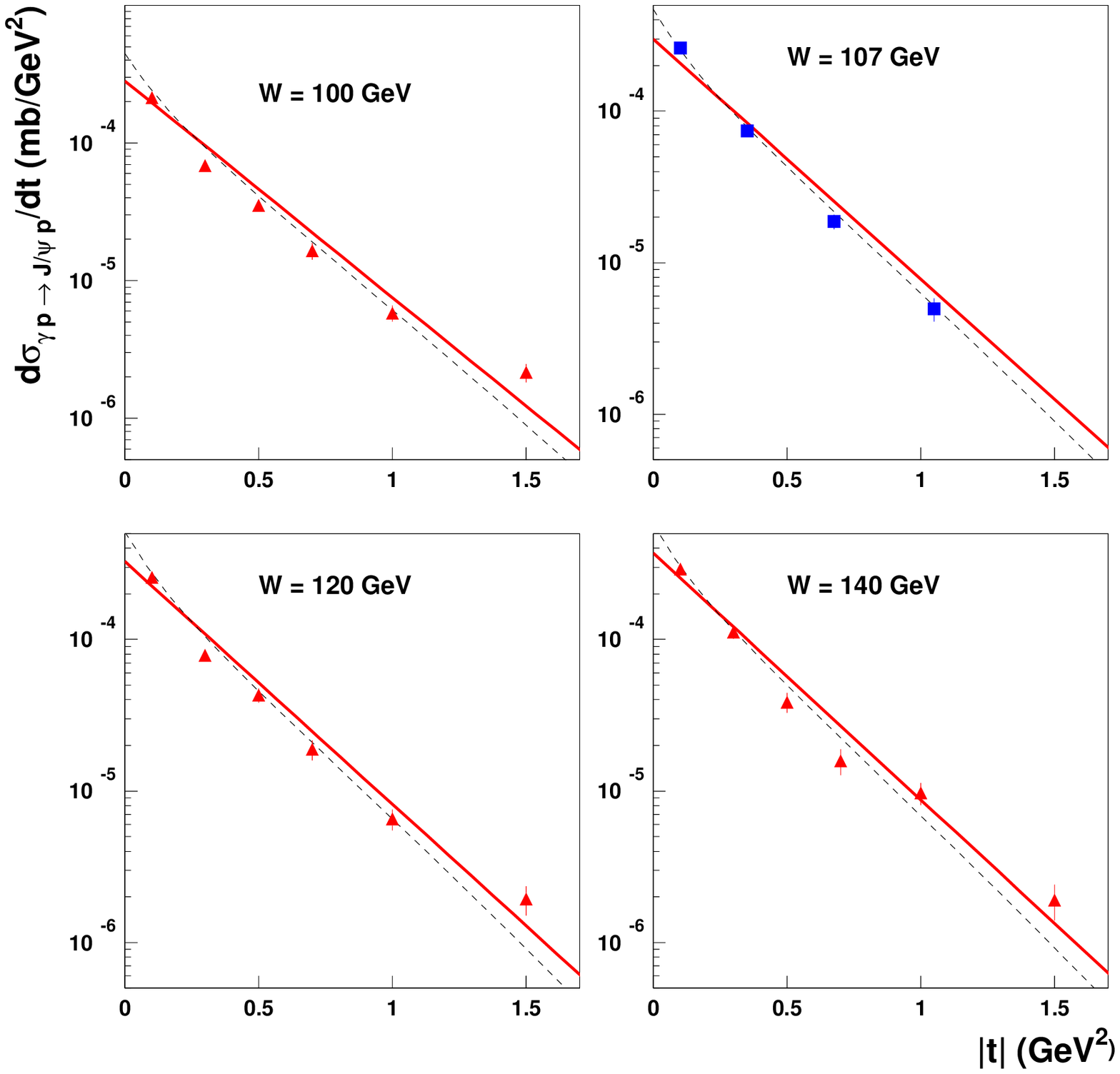}}}
\hskip 1cm
\caption{}
\end{figure}

\newpage

\begin{figure}[hb]
\vskip 1.5cm
\vbox to 18cm {\hbox to 16cm{\epsfxsize=16cm\epsfysize=18cm\epsffile{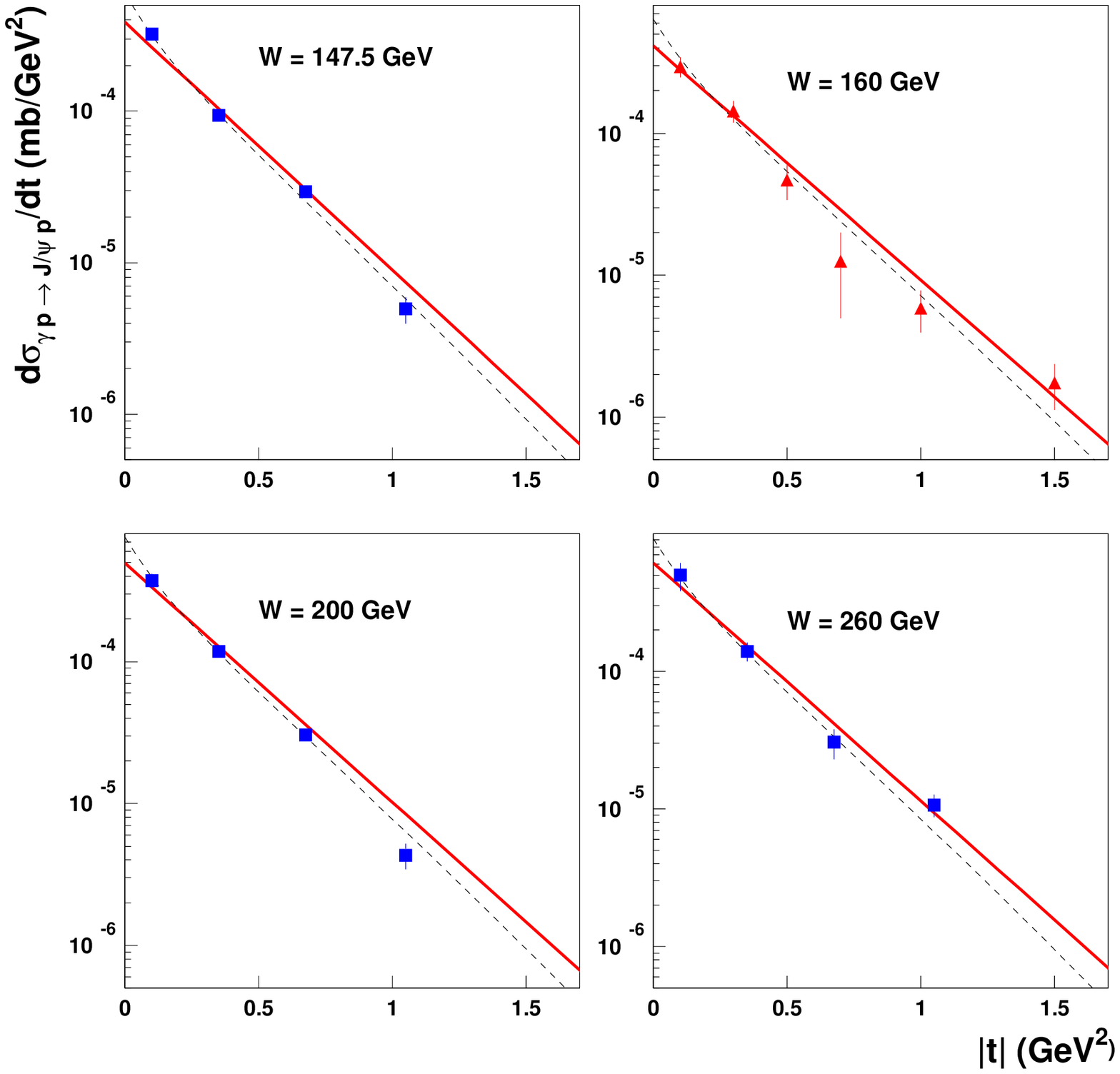}}}
\hskip 1cm
\caption{}
\end{figure}

\newpage

\begin{figure}[hb]
\vskip 1.5cm
\vbox to 18cm {\hbox to 16cm{\epsfxsize=16cm\epsfysize=18cm\epsffile{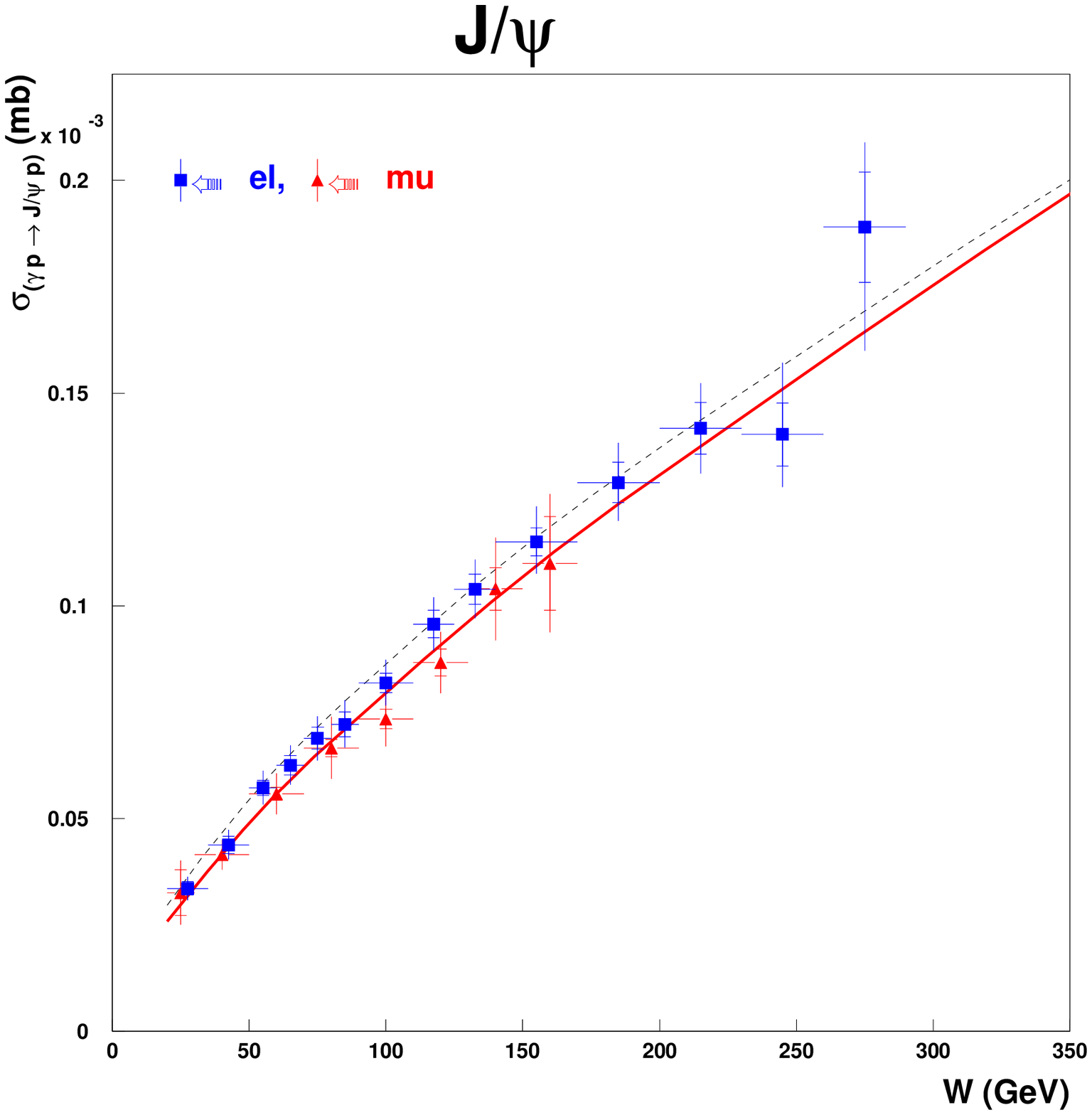}}}
\hskip 1cm
\caption{}
\end{figure}

\end{document}